\documentclass[twocolumn,showpacs,amsmath,amssymb,pre]{revtex4}

\usepackage{graphicx}
\usepackage{dcolumn}
\usepackage{bm}
\usepackage{amsmath}
\usepackage{nicefrac}

\begin{document}

\title{Experimental investigation of the initial regime in fingering
electrodeposition: dispersion relation and velocity measurements }
\author{Matthias Schr\"oter}
\email{matthias.schroeter@physik.uni-magdeburg.de}
\author{Klaus Kassner} 
\affiliation{Fakult\"at f\"ur Naturwissenschaft, Otto-von-Guericke Universit\"at Magdeburg, 
             Postfach 4120, D-39016 Magdeburg, Germany}

\author{Ingo Rehberg} 
\affiliation{Physikalisches Institut, Universit\"at Bayreuth, D-95440 Bayreuth, Germany}

\author{Josep Claret} 
\author{Francesc Sagu\'es}
\affiliation{Departament de Qu\'\i mica-F\'\i sica, Universitat de Barcelona,
             Mart\'\i  \, i Franqu\`es 1, E-08028 Barcelona, Spain}

\date{\today}

\begin{abstract}
Recently  a fingering morphology, resembling the hydrodynamic Saffman-Taylor instability,
was identified in the 
quasi-two-dimensional electrodeposition of copper.
We present here measurements of the dispersion relation of the growing front.
The instability is accompanied by gravity-driven convection rolls
at the electrodes, which are examined using particle image velocimetry. 
While at the anode the theory  presented by Chazalviel et al. 
describes the convection roll, the flow field at the cathode is more 
complicated because of the growing deposit. 
In particular, the analysis of the orientation of the velocity vectors 
reveals some lag of the development of the convection roll compared
to the finger envelope.
\end{abstract}


\maketitle

\section{Introduction}
It is sometimes believed, that all interesting phenomena in the universe
happen at interfaces \cite{priv_com}. Following this line of thought, we believe that
the study of the dynamics of interfaces provides a key for understanding
generic features of non-equilibrium phenomena.
The electrochemical deposition of metals from aqueous solutions in
quasi-two-dimensional geometries is an easily accessible 
growth phenomenon of such an interface.
The emerging structures show a broad variety of growth patterns
including fractals, seaweed or dendrites.
For a recent review see \cite{sagues:00} and references therein.

\begin{figure}[htbp]
  \begin{center}
    \includegraphics[angle=-90,width=8.6cm]{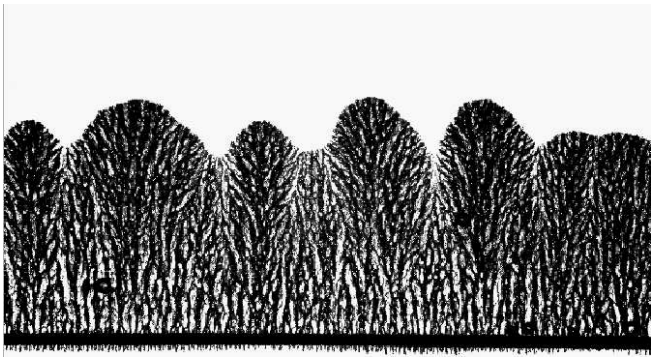}
    \caption{The Finger morphology 403 s after the start of the experiment.
             The solid black line at the bottom of the image is the cathode.
             The  width of the image corresponds to 22.1 mm.
             The deposit was grown in a cell of 250 $\mu$m thickness, the applied
             potential was 15 V.
             The electrolyte contained 50 mM $\mathrm{CuSO_4}$ and 4 mM $\mathrm{Na_2SO_4}$. }
    \label{fig:finger}
  \end{center}
\end{figure}

The focus of this paper is on the electrodeposition of finger deposits 
\cite{trigueros:94}: after the addition of a small
amount of an inert electrolyte like sodium sulphate to a copper sulphate
solution, the morphology of copper deposits changes from a typical fractal
or dense-branched red copper structure to some fine-meshed texture with a
fingerlike envelope. Figure \ref{fig:finger} gives an example of the early
stage of a deposit formed under these circumstances.
The underlying mechanism is believed to be qualitatively understood.
The increase of the
electric conductivity enables alternative reaction paths like the
reduction of $\mathrm{H_{2}0}$. The resulting increase of the $p$H value
triggers the formation of a copper hydroxide gel 
($\mathrm{Cu}_m(\mathrm{OH})_n^{\;(2m-n)+}$) in front of the
advancing deposit \cite{lopez:96,lopez:97}. 
When considering, that the fluid between the copper filaments contains no gel,
the ensuing situation
resembles the Saffman-Taylor instability, where a more viscous
fluid is pushed by a less viscous one and their interface develops the same
type of fingering (see Ref.~\cite{mccloud:95} for a recent survey).

The Saffman-Taylor instability is strongly influenced by the surface tension 
of the interface. In this paper we use that idea to measure the strength
of an effective surface tension associated to the hydrogel-water interface
by analyzing the dispersion relation.
It should be remarked, that the nature of surface tension
between miscible fluids is still an active area of research \cite{fernandez:01}. 

Another necessary ingredient for the occurrence of fingers 
are density-driven convective currents in front of the growing deposit:
If convection is suppressed by turning the 
electrodeposition cell in a vertical configuration,
fingers are not longer formed \cite{lopez:96,lopez:97}. For
that reason it appears to be essential to understand the nature of the
convection field in our experiment, which we 
examine using particle image velocimetry (PIV).

The organization of the paper is the following:
In Section \ref{ch:setup} we introduce the experimental setups. 
Section \ref{ch:disp} is devoted to the dispersion relations, with  \ref{ch:disp_img}
covering some technical aspects and \ref{ch:disp_results}
presenting the measured dispersion relations and analyzing the results
for a textured electrode. In Section \ref{ch:piv}
we discuss the PIV measurements: \ref{ch:piv_result_anode} 
is devoted to the anode while
\ref{ch:piv_result_cat} summarizes our results for the  cathode. 
Finally, Section \ref{ch:conclusion} contains our conclusions.

\section{Experimental Setups}
\label{ch:setup}
\begin{figure}[htbp]
  \begin{center}
    \includegraphics[width=8.6cm]{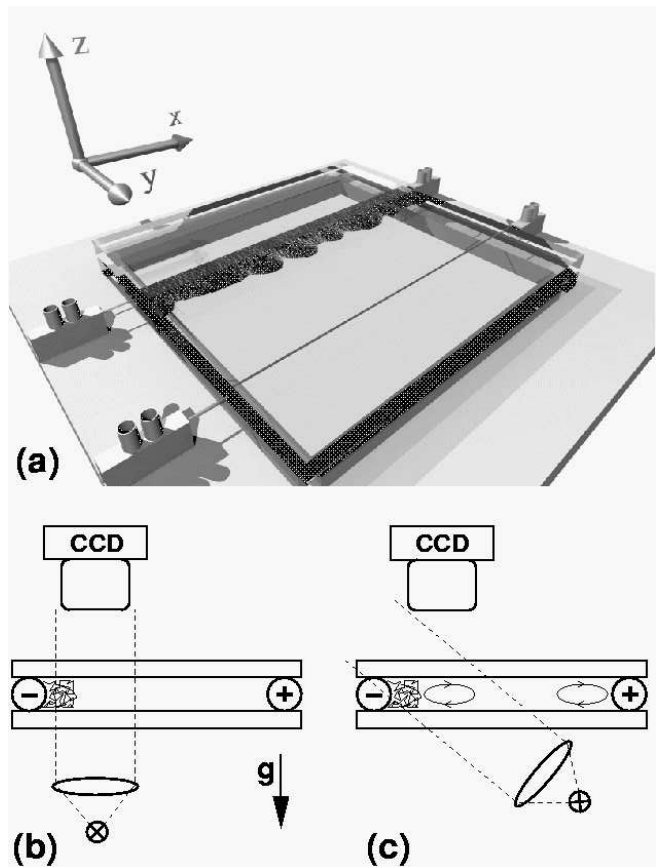}
    \caption{(a) Electrodeposition cell with finger deposit and our choice of the 
      coordinate system. (b) Side view of the illumination with shining-through light 
      used for the measurements of the dispersion relation. Image \ref{fig:finger} was taken
      that way. (c) Side view of the dark field microscopy used for the PIV.
      Image \ref{fig:dark_field} gives an example of this illumination technique.
      The ovals inside the cell indicate the convection rolls.
      }
    \label{fig:setup}
  \end{center}
\end{figure}

The electrodeposition is performed in a cell with two glass plates of 8 x 8 
$\mathrm {cm^2}$ area as side-walls. 
Two parallel copper wires (99.9 \%, Goodfellow) separated by  a distance of 4 cm serve as 
electrodes and spacers. 
Their  diameter $d$ ranges between 125 $\mu$m and 300 $\mu$m.
Figure \ref{fig:setup} a) shows a sketch of this setup. We use
a coordinate system where the x axis is parallel to the electrode, the y axis
points from the cathode to the anode and the z axis is perpendicular to the glass plates.

The space between the electrodes is filled with an aqueous solution of 
50 mM $\mathrm{CuSO_4}$ and 4 mM $\mathrm{Na_2SO_4}$ 
for the measurements of the dispersion relations and 50 mM $\mathrm{CuSO_4}$ and
7 mM $\mathrm{Na_2SO_4}$ for the PIV experiments. All solutions are
prepared from Merck p.a. chemicals in non deaerated ultrapure $\mathrm{H_2O}$.

All measurements are performed with constant (within 0.4 \%) potential between 
the electrodes ranging between 12 V and 19 V. The average current density is below 
35 $\mathrm{mA/cm^2}$.

\begin{table}
  \begin{center}
    \begin{tabular}{l|l|l|l|l|l}
                 &  CCD-camera  & Pixel    &  optical system & resolution & $\Delta t_{aq}$\\ [0.8ex] 
      \hline
       DR        & Kodak          &x: 3070  &  Nikkor 105/2.8  &  7.9 $\mu$m & 5 s    \\
                 & Megaplus 6.3i  &y: 2048  &  SLR lens    &\\
      \hline
       PIV       & Sony          &x: 512  & Olympus  SZH    &  17 $\mu$m  & 2 s   \\
                 & XC 77RR CE    &y: 512  & microscope     &          \\
      \hline
    \end{tabular}

    \caption{Image acquisition systems used in the experimental setups.}  
    \label{tab:setup}
  \end{center}
\end{table}

The two targets of our investigation require two different ways of illumination (sketched in 
Fig.~\ref{fig:setup} (b) and (c)) and image acquisition (summarized in table \ref{tab:setup}).
Since the measurements of the dispersion relation demand a high spatial resolution we 
used a Kodak Megaplus 6.3i CCD-camera
with 3070 x 2048 pixel mounted on a Nikkor SLR macro lens
with 105 mm focal length and a spacer ring.  
To take full advantage of the spatial resolution of 7.9 $\mu$m per pixel it was necessary
to employ a K\"ohler illumination \cite{goeke:88} using filtered  light with a wave 
length of 405 nm from a tungsten lamp.
Images are taken in intervals $\Delta t_{aq}$ of 5 s and are directly transferred with a 
frame-grabber card to the hard disk of a PC.

\begin{figure}[htbp]
  \begin{center}
    \includegraphics[width=8cm,angle=-90]{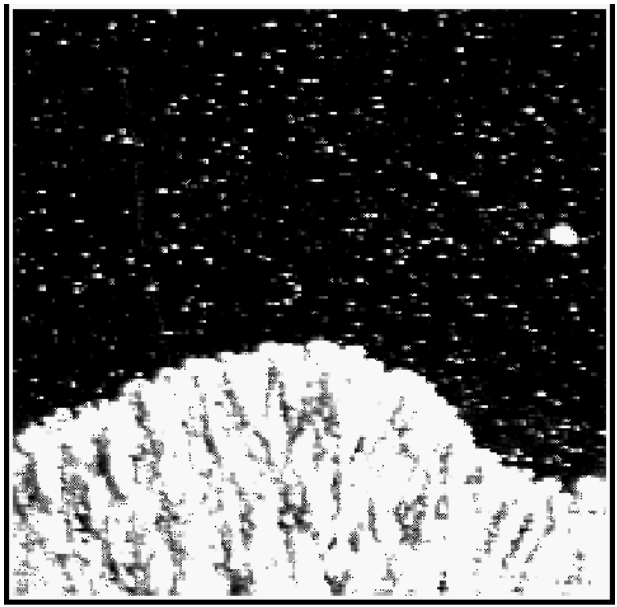}
    \caption{Part of an image taken in dark-field microscopy. Only light scattered 
      from the deposit or the tracer particles is visible. The width corresponds
      to 3.4 mm. (Cell thickness: 300 $\mu$m, applied potential: 12 V,
      solution: 50 mM $\mathrm{CuSO_4}$ and 7 mM $\mathrm{Na_2SO_4}$.) }
    \label{fig:dark_field}
  \end{center}
\end{figure}

To visualize the velocity field inside the cell, we added latex tracer particles to 
the electrolyte. 
We used particles with 0.3 $\mu$m diameter, which stay suspended due to Brownian motion.
Since we cannot resolve these particles with our optical system, we used 
dark-field microscopy: only light scattered from objects inside the cell falls into the lens. 
Image \ref{fig:dark_field} gives an example, the white area at the bottom represents the growing
deposit, the points above correspond to tracer particles.

We did not observe electro-osmosis as reported in \cite{huth:95},
but the particles show some tendency to coagulate and settle to the bottom plate. 
This problem is handled in later stages of the image processing.

Images were acquired using an Olympus SZH stereo microscope and a Sony XC 77RR CE
CCD-camera with 512 x 512 pixels which resulted in a spatial resolution of 17 $\mu$m per pixel.
$\Delta t_{aq}$ was 2 s and images were also directly transferred to a PC.
 
\section{Dispersion relation}
 \label{ch:disp}
To characterize  the instability of a pattern forming systems there is a
quite common method: starting with the uniform system, one adds a small
sinusoidal perturbation of wavenumber $k$ and amplitude $A_{0}$ and
investigates its temporal evolution. As long as the system is
in the linear regime the perturbation will grow or shrink exponentially: 
\begin{equation}
  \label{eq:exp_growth}
  A(k, t)=A_{0}\;e^{\sigma (k)\;t}  
\end{equation}
The dependency of the growth rate $\sigma $ on the wavenumber $k$ is
called dispersion relation.

In the context of electrodeposition dispersion relations have been measured
for the initial phase of compact  \cite{kahanda:92} and ramified \cite{bruyn:96} growth
and calculated  to explain the stability of the dense radial morphology
\cite{grier:93}.

\subsection{Image processing}
 \label{ch:disp_img}
\begin{figure}[htbp]
  \begin{center}
    \includegraphics[width=8.6cm]{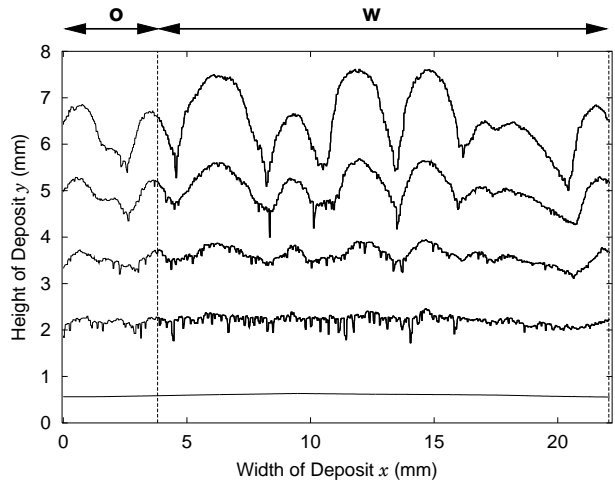}
    \caption{Result  $h(x,t)$ of the front tracking. The curve at the bottom corresponds to the 
      initial position of the cathode. Above the front at 156 s, 259 s, 362 s and 466 s
      after the beginning of the experiment. The experiment is the same as the one 
      presented in Fig.~\ref{fig:finger}.
      The part of length $w$ between the dashed 
      lines was chosen by the cut-out algorithm for further processing.}
    \label{fig:front_dev}
  \end{center}
\end{figure}

In order to measure the dispersion relation
the first step is to track down the temporal evolution of the front:
in the image taken at time $t$ we identify for each column $x$ the height $h(x,t)$
of the deposit.
This is done in two steps:
First we search  for the pair of pixels with the highest gray value gradient in each column.
Second we perform a sub-pixel interpolation by calculating the point of inflexion between
these two pixels. By evaluating a stationary edge, we could demonstrate
that the error of this algorithm is smaller than 0.1 pixel. However the random fine structure
of the deposit acts like the addition of shot noise to $h(x,t)$. We try to mitigate this effect
by applying a median filter with 7 pixel width
\cite{jaehne:97}. Figure \ref{fig:front_dev} shows $h(x,t)$ for
4 various times of an experiment.

Our next aim is to do a Fourier decomposition of $h(x,t)$ to examine the dynamics of the  
amplitudes $A(t,m)$ of the individual modes $m$.
Because our initial conditions are random noise, which can be understood as a superposition
of various modes with different wavelengths $\lambda$, we encounter a leakage problem:
We observe our system in a window of width $w_0$. The ratio $w_0 / \lambda$ 
of the dominant modes is almost always fractional. 
This has a very unpleasant consequence for the Fourier transformation,
which computes the amplitudes for integer mode numbers:
power of the dominant modes is transfered to other less excited modes and 
spoils the measurement of the growth rates there. 

A common answer to this issue is the usage of a windowing function \cite{harris:78},
which does not eliminate leakage, but restricts it to neighboring mode numbers.
The price to be paid is, that this leakage will now occur even if $w_0$ 
is a common multiple of all wavelengths contained in the initial signal. 

In this paper we use a different approach, which is illustrated in Fig.~\ref{fig:front_dev}:
we cut out a part of $h(x,t)$ of width $w$ and offset $o$ 
and use only this part  for the subsequent analysis.
$o$ and $w$, which are constant for the whole run of the 
experiment, are chosen such 
that the left and the right end of the cut-out have the same height:
\begin{equation}
  \label{eq:rwod_1}
  h(o,t) - h(o+w,t) = 0
\end{equation}
and the same slope:
\begin{equation}
  \label{eq:rwod_2}
  \frac{\partial h(o,t)}{\partial x} - \frac{\partial h(o+w,t)}{\partial x} = 0
\end{equation}
Fulfilling Eq.~\ref{eq:rwod_1}  and Eq.~\ref{eq:rwod_2} for all times 
is for all practical purposes identical to the statement that  $w$ is 
a common multiple of all wavelengths contained in $h(x,t)$, that is
the condition under which no leakage occurs.

In practice,  Eq.~\ref{eq:rwod_1} and Eq.~\ref{eq:rwod_2} can be satisfied
only approximately. The algorithm followed to select
$o$ and $w$ minimizes the sum over the height differences according to
Eq.~\ref{eq:rwod_1} while it assures, that the sum over the slope differences
according to Eq.~\ref{eq:rwod_2} does not exceed a threshold.
The average height difference obtained in that way is $\le$ 3 pixels
while the slope difference threshold was 1 pixel/pixel, 
both values correspond to the noise level of $h(x,t)$.
The final step is a standard Fourier decomposition of the cut-out parts of $h(x,t)$.

\subsection{Results}
\label{ch:disp_results}
\begin{figure}[htbp]
  \begin{center}
    \includegraphics[angle=-90,width=8.6cm]{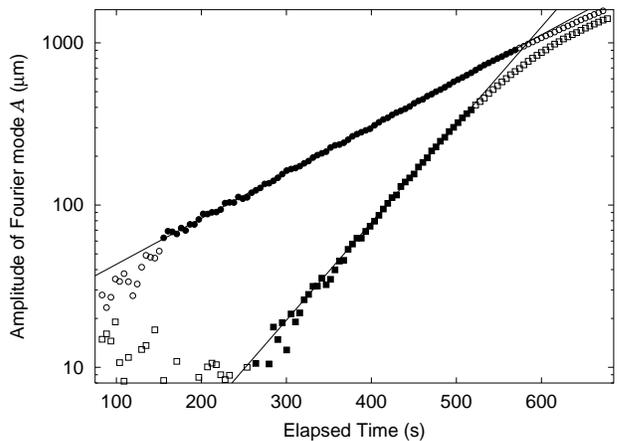}
    \caption{Exponential growth of the Fourier modes.
      The wave numbers are 6.9 $\mathrm{cm^{-1}}$ (circles) and  13.8  $\mathrm{cm^{-1}}$ 
      (squares).
      Only the data points represented by filled symbols were included in the
      fits with Eq.~\ref{eq:exp_growth}. 
      The experiment is the same as in Fig.~\ref{fig:front_dev}.}
    \label{fig:wachstum}
  \end{center}
\end{figure}

For each Fourier amplitude we tried to fit the temporal evolution with
the exponential growth law given by Eq.~\ref{eq:exp_growth}. Fig.~\ref{fig:wachstum} gives an
example for two modes, corresponding to the experiment displayed in Fig.~\ref{fig:front_dev}.
The time interval for the exponential fits is indicated by the solid symbols. 
The start time is set by the time the hydrogel layer needs to build up; 
before that point no instability or destabilization of the growing interface is clearly evidenced.
In the experiment shown in Fig.~\ref{fig:front_dev} this point corresponds to
an elapsed time of about 150 s.   
However some modes need longer until they are grown to an amplitude which 
can be distinguished from the measurement background noise. 
At the other end the fit is limited by the onset of deviations from the exponential
growth law due to nonlinear effects becoming important.

\begin{figure}[htbp]
  \begin{center}
    \includegraphics[angle=-90,width=8.6cm]{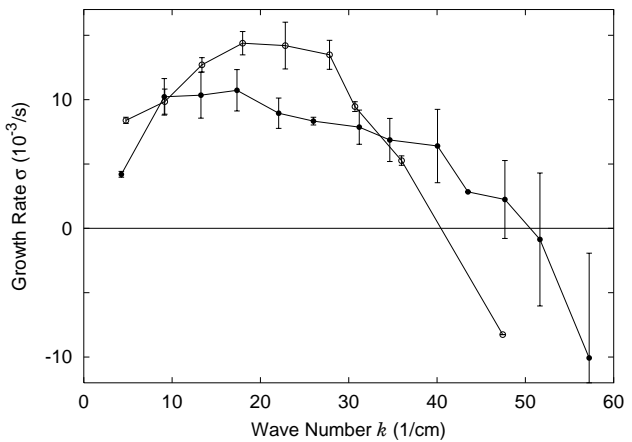}
    \caption{Dependence of the dispersion relation on the applied potential.
      Filled circles correspond to experiments performed applying 15 V, open circles
      applying 19 V. Cell thickness was 250 $\mu$m in both cases.
      All data sets are averaged over three experiments,
      error bars give the standard deviation of the mean value.}
    \label{fig:disp_u}
  \end{center}
\end{figure}

To test the dependence of the dispersion relation on the applied potential,
we performed measurements with 15 V and 19 V using electrodes of 250 $\mu$m 
diameter. The results are displayed in Fig~\ref{fig:disp_u}, each is averaged 
over three experiments.
Both dispersion relations show a limited band of positive 
amplitude growth rates with $k$ between zero and $k_{\mathrm{crit}}$ 
and a wave number $k_{\mathrm{max}}$ where the growth rate is maximal. 

The increase of $U$ from  15 V to 19 V is accompanied by an increase 
of the average growth velocity $v$ from 13.3 $\mu$m/s to 15.9  $\mu$m/s.
The distinct shift of $k_{\mathrm{max}}$ to higher wave numbers is
in conformity with measurements of the number of incipient fingers as a 
function of $v$ reported in \cite{lopez:96}. The decrease of 
$k_{\mathrm{crit}}$ could originate from a change of the physicochemical properties
of the copper hydrogel.

\begin{figure}[htbp]
  \begin{center}
    \includegraphics[angle=-90,width=8.6cm]{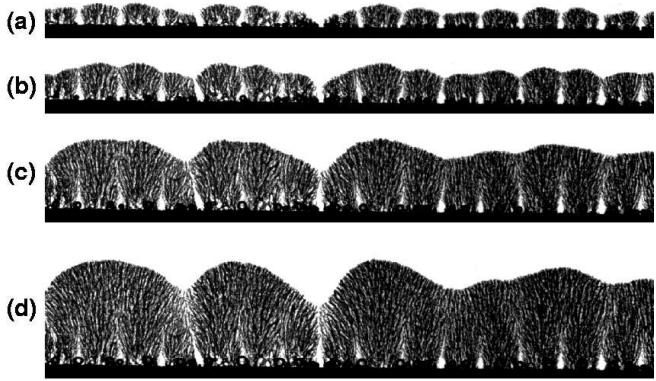}
    \caption{Growing deposit at the textured electrode.
      Images were taken (a) 51 s, (b) 82 s, (c) 142 s, (d) 202s
      after the beginning of the experiment. The width of the images
      corresponds to 15.8 mm.
      (Applied potential: 15 V,
      cell thickness 155 $\mu$m,
      solution: 50 mM $\mathrm{CuSO_4}$ and 7 mM $\mathrm{Na_2SO_4}$.) }
    \label{fig:strukt_devel}
  \end{center}
\end{figure}

\begin{figure}[htbp]
  \begin{center}
    \includegraphics[angle=-90,width=8.6cm]{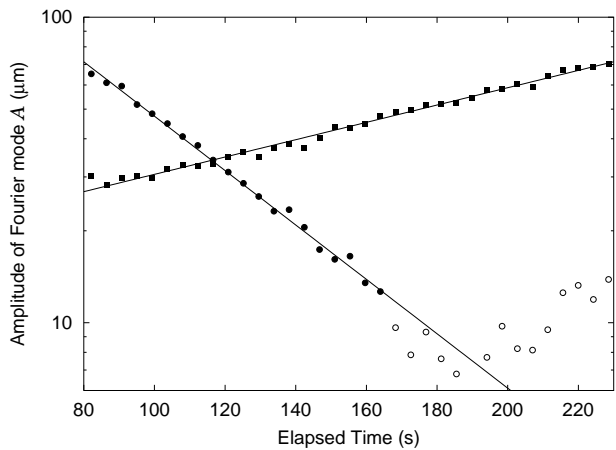}
    \caption{
      Temporal evolution of the Fourier amplitudes measured with the textured
      electrode. 
      The data belong to the experiment displayed in Fig.~\ref{fig:strukt_devel}.
      The wavenumbers are 3.5  $\mathrm{cm^{-1}}$ (squares) and 62.6  $\mathrm{cm^{-1}}$ (circles). 
      Data points represented by filled symbols were included in the fits with Eq.~\ref{eq:exp_growth}.
      }
    \label{fig:strukt_devel_modes}
  \end{center}
\end{figure}

Negative ``growth rates'' can only be measured if the initial 
amplitude of the corresponding mode is strong enough.
To enforce this  we prepared a textured electrode which stimulates 
the initial growth at a wavenumber of 62.8 $\mathrm{cm^{-1}}$. It consists
of a synthetic substrate of 120 $\mu$m height with a 35 $\mu$m copper plating.
The copper layer was etched to derive a comb-like structure with copper stripes of
width 0.75 mm and spacings of 0.25 mm. 
Islands of growing deposit evolve at the tips
of the copper stripes and amalgamate after some time. Fig.~\ref{fig:strukt_devel} 
gives an example. The temporal evolution of two Fourier modes of this experiment is
shown in  Fig.~\ref{fig:strukt_devel_modes}.

\begin{figure}[htbp]
  \begin{center}
    \includegraphics[angle=-90,width=8.6cm]{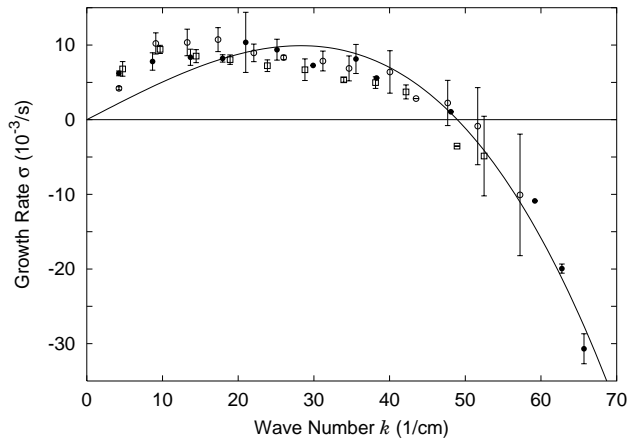}
    \caption{Dependence of the dispersion relation on the cell thickness.
       Open circles were measured in cells with  $d$ = 250 $\mu$m, 
       open squares in cells with $d$ = 125 $\mu$m. The closed circles give 
       the results of the textured electrode ($d$ = 155 $\mu$m). 
       The solid line is a fit of  Eq.~\ref{sata_disprel} to the growth
       rates of the textured electrode.
       Applied potential was 15 V in all cases.
       All data sets are averaged over three experiments,
       error bars give the standard deviation of the mean value.}
    \label{fig:disp_d}
  \end{center}
\end{figure}

The dispersion relation of the textured electrode is displayed in Fig~\ref{fig:disp_d} together
with results for cell thickness of 125 $\mu$m and  250 $\mu$m.
Within the scope of our experimental errors no influence of $d$ is observable.
In principle a decrease of $d$ is accompanied by a decrease of the surface tension forces, 
which should result in a higher  $k_{\mathrm{crit}}$.
However as discussed in chapter~\ref{ch:piv} convection will also be significantly reduced.
This will result in a steeper transition between the hydrogel and the electrolyte, which 
increases the surface tension $\gamma$ \cite{smith:81}. 
Presumably these two effects cancel each other.

While we are not claiming that the Saffman-Taylor dispersion relation:
\begin{equation}
  \sigma = \frac{(\eta_{\mathrm gel} - \eta_{\mathrm H_20}) v}{\eta_{\mathrm gel} + \eta_{\mathrm H_20}} k - 
           \frac{d^2 \gamma}{12 (\eta_{\mathrm gel} + \eta_{\mathrm H_20})} k^3 
  \label{sata_disprel}
\end{equation}
gives a full explanation of this fingering phenomenon, we do believe that it
captures the essentials of the physical mechanism, especially the damping
effects due to an effective surface tension. 
Therefore we performed a fit of Eq.~\ref{sata_disprel}
to the dispersion relation of the textured electrode, which is shown in Fig.~\ref{fig:disp_d}.
As results we find  a viscosity of the hydrogel $\eta_{\mathrm gel}$ of 2.4 $10^{-3}$ kg/ms which is
about twice the viscosity of water $\eta_{\mathrm H_20}$.
The effective surface tension  $\gamma$ turns out to be 3.5 $10^{-7}$ N/m. This is
about five  magnitudes lower than the surface tension at the water-air interface, which is 
presumably due to the fact, that the gel and the water are miscible fluids.

\section{Velocity measurements}
\label{ch:piv}
Electrodeposition is often accompanied by buoyancy-driven convection rolls
\cite{rosso:94,barkey:94,huth:95,linehan:95,chazalviel:96,argoul:96,lopez:97a,dengra:00}. 
The driving force for the convection  are the concentration changes at the electrodes:
at the anode the ion concentration and therefore the density of the 
electrolyte increases. 
While it descends, lighter bulk solution flows in and a convection roll as
sketched in Fig~\ref{fig:setup} (c) starts to grow. 

To visualize the growing deposit in the $x$-$y$ plane, we observe the cells from above.
As apparent from Fig.~\ref{fig:setup} (c) the plane defined by the convection roll is the
$y$-$z$ plane. This results in the uncomfortable situation, that the observed tracer particles 
move simultaneously towards and away from the electrode, which obviates the use of standard
correlation  techniques \cite{raffel:98} for the PIV.

We therefore developed a software package capable to keep track of the motion of 
individual particles. This is done in two steps:
First we identify the particles in each image and insert their center of mass into a database.
Then we construct contiguous histories for individual particles using 
five consecutive time-steps.
In this way we are able to measure the $v_{\rm x}$ and
 $v_{\rm y}$ components of the flow.
The software is published under the GNU Public License and can be downloaded from 
a URL~\cite{download:01}.

\subsection{Anode Results}
\label{ch:piv_result_anode}
\begin{figure}[htbp]
  \begin{center}
    \includegraphics[angle=-90,width=8.6cm]{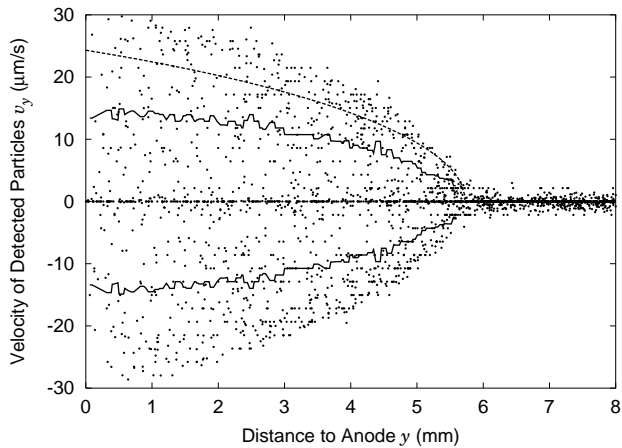}
    \caption{Particle velocities at the anode after 888 s.
      Particles with positive $v_{\mathrm{y}}$ are moving away from, with negative
      $v_{\rm y}$ towards the anode.
      Only particles lying outside the solid lines were considered
      for the calculation of an average velocity $v_{\mathrm{avg}}$.
      The dashed line is a fit of Eq.~\ref{eq:fit_v_y} to
      $v_{\mathrm{avg}}(y,t)$.
      The applied potential is 12 V, the cell thickness 300 $\mu$m.}
    \label{fig:profil_anode}
  \end{center}
\end{figure}

All velocity measurements were performed in cells with thickness $d$ = 300 $\mu$m
applying a potential of 12 V.
At the anode a convection roll is clearly visible: while having no relevant 
velocity component in the $x$-direction, the tracer particles in a distinct zone move
towards or away from the electrode. Fig.~\ref{fig:profil_anode} gives the
$v_{\rm y}$ components of all particles detected in one image as a function of their
distance $y$ to the anode.

Due to the big depth of focus of our optical system, 
we observe particles in all heights $z$ of the cell simultaneously. 
As the particle velocity is a function of $z$, we find 
for a given $y$  all velocities between $\pm  v_{\mathrm{max}}(y,t)$.

\subsubsection{Theory}
It is known that some time after the start of an experiment
vertical diffusion starts to smear out the concentration differences between the flows
to and away from the electrode. Chazalviel and coworkers \cite{chazalviel:96} proposed 
a two-dimensional description for this diffusion-hindered spreading (DHS) regime. 
The velocity component perpendicular to the electrode $v_{\rm y}$ should obey \cite{exchange}:
\begin{eqnarray}
  v_{\rm y}(y,z,t) = && k_1 \; \left[  \left(1 - \frac{y}{k_2 \; \sqrt{t}}\right)^{\nicefrac{1}{2}} 
           - \frac{1}{12}\left(1 - \frac{y}{k_2 \; \sqrt{t}}\right)^{\nicefrac{3}{2}} \right]
         \nonumber\\
         &&     \left( z^3 - z \frac{d^2}{4} \right)   
  \label{eq:velocity_y}
\end{eqnarray}
with
\begin{equation}
  \label{eq:k_1}
   k_1 = 13.1 \left( \frac{i \; \mu_{\mathrm a}  \; \frac{\partial \rho}{\partial c} \; g}{z_{\mathrm c} \; F \; (\mu_{\mathrm a} + \mu_{\mathrm c})\; \eta } \right)^{\nicefrac{1}{3}} \;\frac{D^{\nicefrac{1}{3}}}{ d^{\nicefrac{8}{3}}}
\end{equation}
and
\begin{equation}
  k_2 = 0.222 \left( \frac{i \; \mu_{\mathrm a}  \; \frac{\partial \rho}{\partial c} \; g}{z_{\mathrm c} \; F \; (\mu_{\mathrm a} + \mu_{\mathrm c})\; \eta } \right)^{\nicefrac{1}{3}} \; \frac{d^{\nicefrac{4}{3}}}{D^{\nicefrac{1}{6}}}
\label{eq:k_2}
\end{equation}
$i$ denotes the current density (25 $\pm$ 3 mA/$\mathrm{cm^2}$), 
$\frac{\partial \rho}{\partial c}$ is the dependency of the density on the ion concentration, 
which we measured to be 0.156 $\pm$ 0.008 kg/Mol for $\rm CuSO_4$
using a density measurement instrument DMA 5000 from Anton Paar. 
$\mu_{\rm a}$ and $\mu_{\rm c}$ represent the  mobility of the anions (8.3 $10^{-8}$ $\mathrm m^2$/sV)
and cations (5.6 $10^{-8}$ $\mathrm m^2$/sV), respectively.
$F$ is the Faraday constant (9.6 $10^{4}$ As/Mol), $g$ the acceleration due to gravity and 
$z_{\mathrm c}$ the charge number of the cation (2).
$\eta$ represents the dynamic viscosity ($10^{-3}$ kg/ms) of the solution and 
$D$ the ambipolar diffusion constant
(8.6 $10^{-10}$ $\rm m^2$/s for $\rm CuSO_4$ \cite{crc:94}). Because $D$ and $\eta$
are weakly concentration dependent, we assign them errors of 10 \% in the subsequent calculations,
further on we assume a 5 \% error in the determination of $d$. 
Eq.~\ref{eq:velocity_y} assumes, that the glass plates are at $ z = \pm d/2$ and $y$ = 0 at the anode.

Within this theory the extension $L$ of the convection roll  is given by the point, 
where $v_{\rm y}(y,t)$ drops to zero:
\begin{equation}
  L  = k_2 \: \sqrt{t}
  \label{eq:length_roll}
\end{equation}

The maximal velocity $v_{\rm max}$ occurs at the heights $z = \pm d/2\sqrt{3}$, in the immediate
neighborhood of the electrode $(y=0)$, and is time independent: 
\begin{equation}
  \label{eq:v_max}
  v_{\rm max}= 0.63 \left( \frac{i \; \mu_{\mathrm a}  \; \frac{\partial \rho}{\partial c} \; g\; d \; D}{z_{\mathrm c} \; F \; (\mu_{\mathrm a} + \mu_{\mathrm c})\; \eta } \right)^{\nicefrac{1}{3}}
\end{equation}
However the authors state that due to problems with the boundary conditions at the electrode
their solution might be not applicable in the region $0 < y < d$.

\subsubsection{Test of the theory}
In principle the theory should describe directly our experimental results, however due to the
uncertainties in the parameters we decided to perform a fit. 
Because we measure a two-dimensional projection of velocities, we concentrate on an average velocity
$v_{\rm avg}(y,t)$, which  
we calculate from all particles with absolute velocities of at least half the velocity of the 
fastest particle in this distance. In Fig~\ref{fig:profil_anode} this corresponds to all particles 
not lying between the two solid lines. 
This restriction is necessary to remove the 
contribution of particles which have already settled to the bottom glass plate.

\begin{figure}[htbp]
  \begin{center}
    \includegraphics[angle=-90,width=8.6cm]{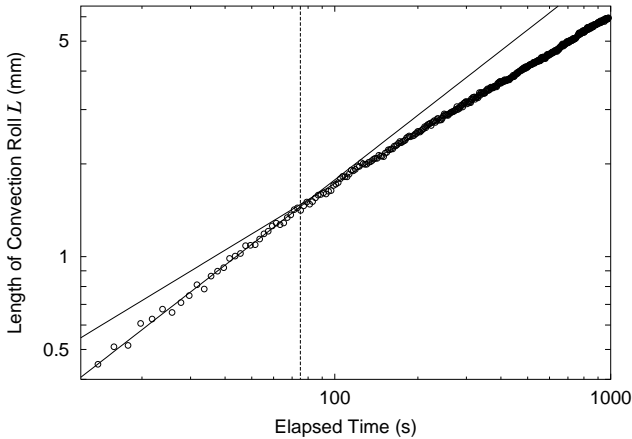}
    \caption{Length $L$ of the convection roll at the anode. $L$ was determined
      by fitting $v_{\mathrm{avg}}(y,t)$ with Eq.~\ref{eq:fit_v_y}. The two fits
      to the left and right of the dashed line 
      have slopes of 0.7 $\pm$ 0.01 and  0.543 $\pm$ 0.001 where the errors are
      the asymptotic standard errors derived from the covariance matrix of the 
      fit \cite{numre:92}.
      The experiment is the same as presented in Fig.~\ref{fig:profil_anode}.}
    \label{fig:rolle_anode}
  \end{center}
\end{figure}

From Eq.~\ref{eq:velocity_y} we derive our fit function:
\begin{equation}
  \label{eq:fit_v_y}
  v_{\mathrm{avg}}(y,t) =  v_0 \left[   \left(1 - \frac{y}{L}\right)^{\nicefrac{1}{2}} 
         - \frac{1}{12}\left(1 - \frac{y}{L}\right)^{\nicefrac{3}{2}} \right]
\end{equation}
The fit of Eq.~\ref{eq:fit_v_y} to $v_{\rm avg}(y,t)$ is successful for the whole run
of the experiment, especially the roll length $L$ is determined very precisely. 
Fig.~\ref{fig:rolle_anode} illustrates the temporal evolution of $L$ obtained from the fit.
After 70 s $v_{0, \mathrm{exp}}$ becomes constant at a value of  24.7 $\pm$ 0.5 $\mu$m/s.
Integrating Eq.~\ref{eq:velocity_y} yields $v_{0, \mathrm{theo}}$ = 30.8 $\pm$ 2.1 $\mu$m/s.

The next step is testing Eq.~\ref{eq:length_roll}, which
predicts the growth law of $L$. A fit of our experimental
$L$ for times greater 75 s using equation $L(t) = k_{2,fit} t^b$
yields a slope of 0.543 $\pm$ 0.001, which is slightly 
above the square root law suggested by Eq.~\ref{eq:length_roll}. 
Higher exponents have also been reported by  other groups: Argoul et al.
\cite{argoul:96} found 0.56 $\pm$ 0.01 and
Dengra et al. \cite{dengra:00} measured 0.54 $\pm$ 0.02.
The coefficient $k_{2,fit}$ was found to be 141 $\pm$ 1 $\mu$m/$\rm s^{\nicefrac{1}{2}}$. 
If we insert our experimental parameters in Eq.~\ref{eq:k_2} we obtain 
$k_2$ = 134 $\pm$ 12 $\mu$m/$\mathrm s^{\nicefrac{1}{2}}$ in agreement with the fit.

Fig.~\ref{fig:vergleich_elektroden} (b) shows the experimental results for the maximal 
fluid velocity, which is located in the vicinity of the anode.
After a sharp rise at the beginning of the experiment follows a slightly inclined plateau.
A linear fit yields a velocity of 25.5  $\pm$ 0.1 $\mu$m/s  for $t$ = 0, which increases 
at a rate of 1 \% per minute. 
Inserting the experimental parameters into Eq.~\ref{eq:v_max} results in 
a constant velocity of 37.1 $\pm$ 2.5 $\mu$m/s.
This discrepancy can be attributed to the unphysical boundary conditions used in the model.

Summarizing it can be stated, that the theory presented in \cite{chazalviel:96}
provides a  qualitatively and semi-quantitatively good description of 
the anodic convection roll in the DHS regime.

\subsubsection{Initial phase of development}
Apart from the DHS regime, Fig.~\ref{fig:rolle_anode} also shows 
a growth law $L \sim t^{0.7}$ for times between 12 s and 75 s.
The exponent of 0.7 $\pm$ 0.01 indicates a faster growth originating
from a different mechanism.
In the very beginning of an experiment
convective fluid transport is faster than the diffusive equalization between the 
copper ion enriched electrolyte
and the bulk electrolyte. Therefore the concentrated electrolyte at the electrode sinks down and 
spreads along the bottom plate without significant mixing. 
In this so called immiscible fluid (IF) regime  the length $L$ of the convection zone 
is expected to grow  with $t^{\nicefrac{4}{5}}$.
This has been shown for gravity currents \cite{chen:80,huppert:82} and was successfully
adapted for electrodeposition \cite{huth:95,chazalviel:96}.
As explained in detail below, the range of applicability of this theory requires in our 
case $L <$ 470 $\mu$m, which is out of our measured range.
Thus we observe a transitional period between the IF and the DHS regime 
and not the IF regime itself.

To distinguish between the IF and the DHS regime a scaling analysis based on the 
vertical diffusion time was used in \cite{huth:95}. 
We would like to advocate a different approach, using the similarity of the driving mechanism
with the well-investigated case of a side heated box filled with fluid
\cite{cormack:74,imberger:74,patterson:80}. 
While in this case the density changes are due to thermal expansion, there exist also 
two flow regimes:
In the so called convective regime small layers of fluid spread along the confining 
plates with a stagnant core in the middle of the cell.
The conductive regime is characterized by a cell filling convection roll and iso-density lines 
which are almost vertical. 
These flows can be described by the Rayleigh number $Ra = \alpha \Delta T \rho g d^3 / \kappa \eta$ 
and the aspect ratio $A = d/W$. $\alpha$ is the thermal expansion and $\kappa$ the thermal
diffusivity of the fluid. $W$ denotes the distance,  
$\Delta T$ the temperature difference between hot and cold side wall. Boehrer \cite{boehrer:97} pointed out, that $Ra \, A^2$,
which equals the ratio between the timescales for vertical diffusion and horizontal convection,
is the dimensionless control parameter of this transition. For high values of $Ra \, A^2$ one
observes the convective regime, for small values the conductive one.

If we transfer this analysis to our situation, we have to substitute the thermal density difference
$\alpha \Delta T \rho$ with the density difference due to concentration changes 
$ \frac{\partial \rho}{\partial c} c_0 $ and the thermal diffusivity $\kappa$ with the ambipolar 
diffusivity $D$. This yields a concentration-dependent Rayleigh number $Ra_{\rm c}$ 
expressed by:
\begin{equation}
  \label{eq:ra_c}
  Ra_{\rm c} = \frac{\frac{\partial \rho}{\partial c}\; c_0 \; g\; d^3 }{D \; \eta}
\end{equation}
The aspect ratio is calculated using the length of the convection roll: $A = d/L$.
Analyzing the results presented in Fig 10 of Ref.~\cite{huth:95}
this interpretation provides a necessary condition
given by:
\begin{equation}
  \label{eq:cond_if}
  Ra_{\rm c} \, A^2 > 1000
\end{equation}
to observe the IF regime. In our experiment $Ra_{\rm c} \, A^2$ will only be larger than 1000  for 
$L <$ 470 $\mu$m , which is out of our measured range.

\subsection{Cathode Results}
\label{ch:piv_result_cat}
\begin{figure}[htbp]
  \begin{center}
    \includegraphics[angle=-90,width=8.6cm]{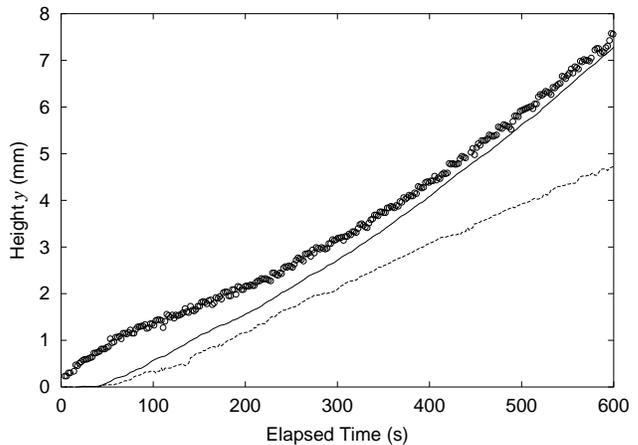}
    \caption{Temporal evolution at the cathode. The solid line describes the position 
      of a finger tip, the dashed line of a neighboring valley. The circles mark the
      leading edge of the convection roll.
      The experimental conditions are: $U$ = 12 V and $d$ = 300 $\mu$m.}
    \label{fig:rolle_katode}
  \end{center}
\end{figure}

At the cathode the situation is more complex due to the growing deposit.
In Fig.~\ref{fig:rolle_katode} the solid line describes the position of the most advanced point
of the deposit in a width of 2.3 mm in $x$-direction, 
while the dotted line corresponds to the minimum in the 
same interval. The border of the convection roll was measured using a threshold: the open
circles mark the foremost position, where the $v_y$ component of a particle exceeded 4.3 $\mu$m/s.

While the convection roll develops immediately, no growth of the deposit is observable in
the first 40 s, because the copper deposits in a planar compact way, which is not observable with
our optical resolution. 
During this so called Sand's time, the ion concentration at the 
cathode drops to zero, which subsequently
destabilizes the planar growth mode \cite{argoul:96}. 

In the next phase (40 s $ < t < 280$ s) 
a depth of the deposit 
(distance between the most advanced and most retarded parts of the growing deposit)
becomes measurable and finally reaches a constant size.
The advancing deposit 
significantly compresses the size $L$ of the convection roll as given by the distance
between the dots and the solid line.
Within this time interval the hydrogel layer is established, which can be seen by visual
inspection. 

The third phase ($t > 280$ s) is characterized by the appearance of the finger development.
The front minimum (dashed line)
and maximum (solid line)
in Fig.~\ref{fig:rolle_katode}  coincide to the finger 
tip and the neighboring valley. The length $L$ of the convection roll in front of the 
finger tip  tends to converge to a constant size,
which has also been observed in the absence of a gel
\cite{huth:95,chazalviel:96,dengra:00}.
The dispersion relations studied in chapter \ref{ch:disp_results} 
were obtained in this phase.

\begin{figure}[htbp]
  \begin{center}
    \includegraphics[angle=-90,width=8.6cm]{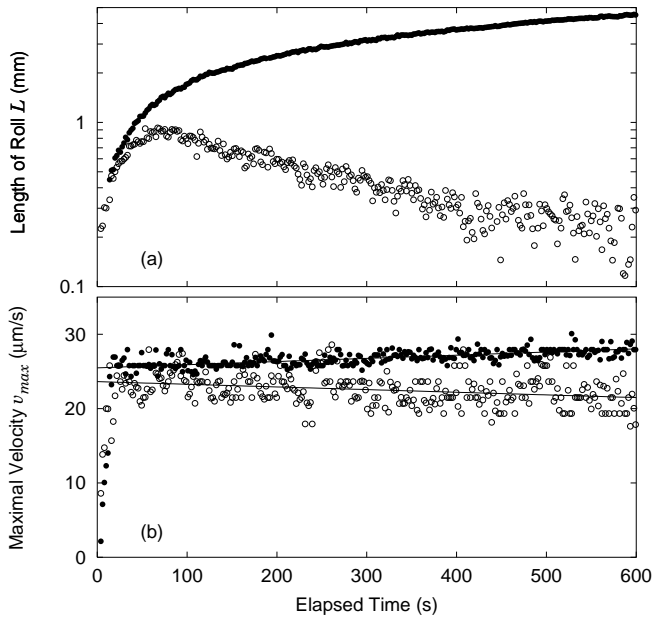}
    \caption{Comparison of the flow behavior at the two electrodes under identical experimental
      conditions. Filled circles correspond to the anode (Fig.~\ref{fig:rolle_anode}), 
      open circles to the cathode (Fig.~\ref{fig:rolle_katode}). 
      Part (a) shows the temporal evolution of the overall
      length $L$ of the convection roll.
      Part (b) gives the temporal evolution of the maximal velocity in y direction
      inside the rolls. The solid lines are fits to all data points with t $>$ 100 s.}
    \label{fig:vergleich_elektroden}
  \end{center}
\end{figure}

Fig.~\ref{fig:vergleich_elektroden} is devoted to a comparison between 
the flow behavior at the two electrodes.
The developments of $L$ displayed in Fig.~\ref{fig:vergleich_elektroden} (a) differ
significantly from each other. 
However the comparison of the maximal fluid velocities in 
Fig.~\ref{fig:vergleich_elektroden} (b) show similarities 
with respect to the absolute value and the approximate temporal constance.
The fit at the cathode yields a velocity of 23.6 $\pm$ 0.3 $\mu$m/s for $t$ = 0 s, 
which decreases 0.9 \% per minute. 

From Fig.~\ref{fig:vergleich_elektroden} (a) we infer, that the theoretical 
description  presented in \cite{chazalviel:96} cannot be applied to the cathode.
Indeed three prerequisites of the theory are not fulfilled:
Most importantly, it does not consider the moving boundary originating from the 
growth process. 
Moreover this model can lead to unphysical negative concentrations at the cathode
due to its inherent simplifications.
Finally the theory is two-dimensional in the $y$-$z$ plane and therefore not able 
to describe the influence of the ramified deposit, which evolves in the $x$-$y$ plane.

\begin{figure}[htbp]
  \begin{center}
    \includegraphics[angle=-90,width=9.5cm]{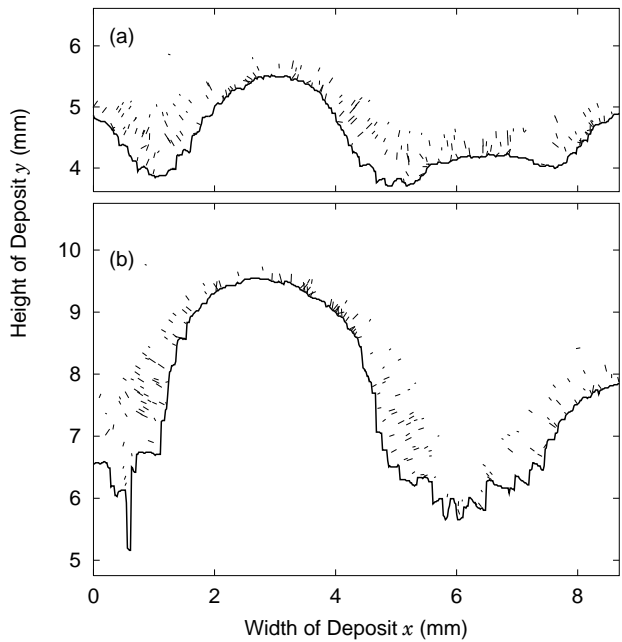}
    \caption{Flow field at (a) 493 s and (b) 760 s after the beginning of 
      the experiment.
      All particles detected in one image with a velocity $>$ 4.8 $\mu$m/s 
      were included. Arrowheads have been omitted for clarity,
      the length of the lines corresponds to the way the particles travel in 6 s.
      ($U$ = 12 V, $d$ = 300 $\mu$m.)}
    \label{fig:felder_stack}
  \end{center}
\end{figure}

Remarkably enough our measurements show that the presence of hydrogel
does not prevent convection.
For a better understanding of the contribution of the flow field to the 
finger morphogenesis, we visualize it in 
Fig.~\ref{fig:felder_stack} for two different times 
of the same experiment. 
The solid line denotes the interface of the deposit, and the arrows indicate
the velocity of individual particles.
It is apparent that convection is restricted to a small zone
in front of the growing deposit.
The  hydrogel occurs also in the immediate vicinity of the front,
however its extension could not be investigated in any detail here.

\begin{figure}[htbp]
  \begin{center}
    \includegraphics[width=8.6cm]{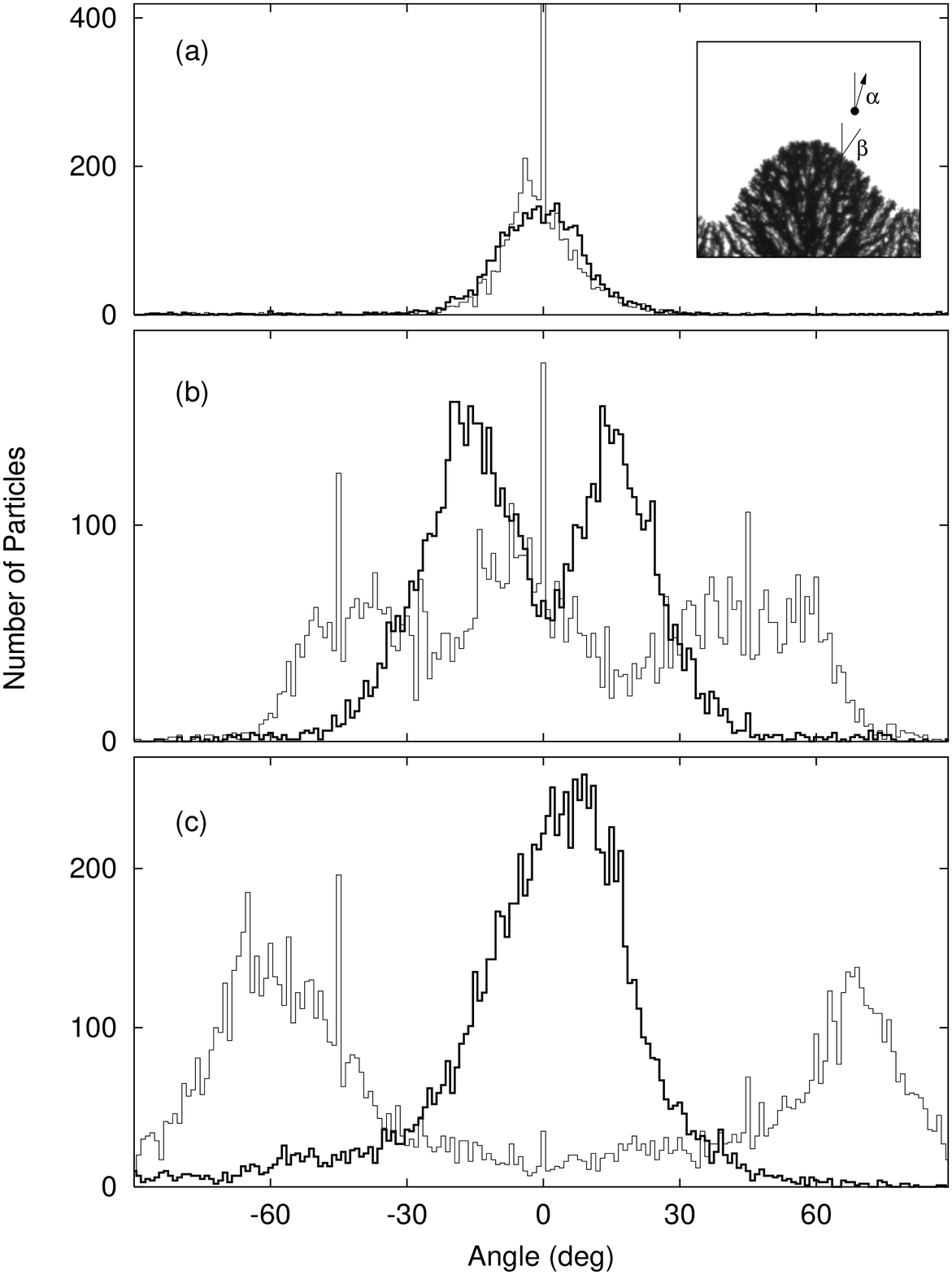}
    \caption{Angle histograms for the times (a) 3-97 s, (b) 493-651 s and (c) 730-918 s.
      The thin line gives the distribution of the angle $\alpha$ between the velocity vector 
      $\mathbf v$
      of a particle and the $y$ axis. Prolonging $\mathbf v$ we determine the meeting point with
      the finger surface. Its normal vector encloses the angle $\beta$ with the $y$ axis.
      The thick line shows the distribution of the angle difference $\alpha - \beta$.
      These data correspond to the experiment presented in Fig.~\ref{fig:felder_stack}.
      }
    \label{fig:winkel_stack}
  \end{center}
\end{figure}

To examine the orientation of the flow field with respect to the interface
we first computed
the distribution of the angle $\alpha$ between the velocity vectors $\mathbf v_i$ of the particles
and the $y$ axis. These data correspond to the thin lines shown in Fig.~\ref{fig:winkel_stack}.
The distribution clearly broadens with time and exhibits two distinct maxima for the fully
developed fingers analyzed in Fig.~\ref{fig:winkel_stack} (c).

Then we identified for each particle an associated point of the finger envelope by prolonging $\mathbf v_i$.
The angle between the normal vector at this meeting point and the $y$ axis is labeled $\beta$.
The angle difference $\alpha - \beta$ is a measure for the mismatch between the deposit
and the convection roll and is displayed as the thick line  in Fig.~\ref{fig:winkel_stack}.

Fig.~\ref{fig:winkel_stack} (b) reveals that the convection field has a mismatch of about 15 degrees
to the front during the initial phase of finger development. 
Thus we conclude that the development of the  convection field lags behind
the development of the front.
After the fingers are fully developed, the convection field
readjusts again perpendicularly to the interface as shown in Fig.~\ref{fig:winkel_stack} (c),
which is a sign of the concentration gradient adapting to the geometry of the deposit. 

Due to the lack of comparable measurements of other electrodeposition
systems we can not judge if this effect is due to hydrogel or a generic
feature of the cathodic convection roll.

\section{Summary and conclusions}
\label{ch:conclusion}
We measured the dispersion relation in the linear
regime of the finger morphology and their dependence on cell thickness and
applied potential. By means of a textured electrode, we were
able to measure negative growth rates. 
The striking feature of the smooth finger envelope is connected with the existence
of a limited band of wavenumber between 0 an $k_{\mathrm{crit}}$ with  positive growth rates. The damping
of all perturbations with higher wavenumbers can be attributed
to an effective  surface tension associated with a hydrogel boundary in front of the deposit
A fit of the dispersion relation yields some  estimates for the effective surface tension and the viscosity 
of the copper hydrogel.

Furthermore we performed PIV measurement at both electrodes. 
At the anode  we could confirm the
$t^{\nicefrac{1}{2}}$ growth law for the length $L$ of the  convection roll.
We determined $L$ by fitting the suggested analytic expression  
for the velocity field and could therefore successfully test the model proposed by 
Chazalviel and coworkers.

At the cathode the maximal fluid velocity inside the convection roll is of the 
same order as at the anode, but the temporal evolution of $L$ differs strongly. 
An analysis of the orientation of the velocity vectors reveals the existence 
of some mismatch between the development of the convection roll and the deposit 
front while the system is in the linear regime.
In the fully developed finger regime, the velocity vectors are again 
perpendicular to the envelope of the deposit.

Our results provide a reasonable explanation  why in the absence of gravity-driven convection rolls 
the fingering instability cannot be observed: without convective mixing, 
the concentration gradient at the hydrogel interface can be assumed to be steeper,
which will increases the effective surface tension. In consequence $k_{\mathrm{crit}}$ shifts to 
smaller wavenumbers and the overall growth rates decrease, which will suppress the evolution 
of fingers. 

An alternative explanation assumes two zones within the hydrogel layer.
One part of the hydrogel in immediate vicinity of the deposit will be mixed by the
convection roll and the consequential shear thinning will decrease its viscosity.
In front of it there is a zone of quiescent hydrogel of higher viscosity, 
at the interface between this two the instability takes place. 
In this scenario $k_{\mathrm{crit}}$ is determined by the length of the convection roll $L$.

A fully quantitative theoretical analysis remains to be done. 

\begin{acknowledgments}
We want to  thank  Marta-Queralt L\`opez-Salvans, Thomas Mahr,
Wolfgang Sch\"opf, Bertram Boehrer, Ralf Stannarius and Peter Kohlert
for clarifying discussions.
We are also indebted to Niels Hoppe and Gerrit Sch\"onfelder, which were instrumental in 
the density measurements and  J\"org Reinmuth  for preparing  the textured electrodes.
This work was supported by the {\it Deutsche Forschungsgemeinschaft} under the projects
En 278/2-1 and FOR 301/2-1. Cooperation was facilitated by the TMR Research Network FMRX-CT96-0085:  
Patterns, Noise \& Chaos.
\end{acknowledgments}


\begin{thebibliography}{33}
\bibitem[{exc()}]{exchange}
\bibinfo{note}{The anion migration velocity $v_{\mathrm a}$ used in
  Ref~\cite{chazalviel:96} is difficult to access experimentally. Therefore we
  did substitute it by the currrent density $i$ using Eq. 5 and Eq. 8 in
  Ref~\cite{chazalviel:96}: $v_{\mathrm a} c_0 = i \mu_{\mathrm a} / z_{\mathrm
  c} F (\mu_{\mathrm a} + \mu_{\mathrm c})$.}
\end{thebibliography}

\end{document}